\documentclass[aps,prl,twocolumn,superscriptaddress]{revtex4-1}
\usepackage{graphicx}

\begin{document}

\title{Proton Kinetic Effects and Turbulent Energy Cascade Rate in the Solar Wind}

\author{K.T. Osman}
\email{k.t.osman@warwick.ac.uk}
\affiliation{Centre for Fusion, Space and Astrophysics; University of Warwick, Coventry, CV4 7AL, United Kingdom}

\author{W.H. Matthaeus}
\affiliation{Bartol Research Institute, Department of Physics and Astronomy, University of Delaware, Delaware 19716, USA}

\author{K.H. Kiyani}
\affiliation{Laboratoire de Physiques de Plasmas, CNRS-Ecole Polytechnique-UPMC, Route de Sacay, 92120 Palaiseau, France}
\affiliation{Centre for Fusion, Space and Astrophysics; University of Warwick, Coventry, CV4 7AL, United Kingdom}

\author{B. Hnat}
\author{S.C. Chapman}
\affiliation{Centre for Fusion, Space and Astrophysics; University of Warwick, Coventry, CV4 7AL, United Kingdom}

\date{\today}

\begin{abstract}

The first observed connection between kinetic instabilities driven by proton temperature anisotropy and estimated energy cascade rates in the turbulent solar wind is reported using measurements from the Wind spacecraft at 1 AU. We find enhanced cascade rates are concentrated along the boundaries of the ($\beta_{\parallel}$, $T_{\perp}/T_{\parallel}$)-plane, which includes regions theoretically unstable to the mirror and firehose instabilities. A strong correlation is observed between the estimated cascade rate and kinetic effects such as temperature anisotropy and plasma heating, resulting in protons 5--6 times hotter and 70--90\% more anisotropic than under typical isotropic plasma conditions. These results offer new insights into kinetic processes in a turbulent regime.  

\end{abstract}

\pacs{}

\maketitle

\textit{Introduction}.---The kinetic processes arising from non-thermal velocity distribution functions (VDF) can affect the macroscopic evolution of space plasmas \citep[see][for reviews]{Marsch06, MatteiniEA12}. In the solar wind, where collisional effects are often weak, particles can exhibit anisotropic distributions with respect to the local magnetic field direction \citep{MarschEA82} such that $R \equiv T_{\perp}/T_{\parallel} \neq 1$. This temperature anisotropy $R$ is dependent on the proton parallel plasma beta $\beta_{\parallel} = n_{p}k_{B}T_{\parallel}/(B^{2}/2\mu_{0})$, which is the ratio of parallel pressure to total magnetic pressure. These parameters do not assume arbitrary values in the solar wind, and \textit{in-situ} measurements suggest proton temperature anisotropy-driven kinetic instabilities are responsible for constraining the plasma \citep{KasperEA02, HellingerEA06, MarschEA06, MatteiniEA07}. From linear Vlasov theory, the solar wind plasma can become unstable to the cyclotron and mirror instabilities when $R > 1$, and to the firehose instability when $R < 1$ and $\beta_{\parallel} \geq 1$. These microinstabilities generate fluctuations that scatter particles towards more isotropic states, and thus could limit the degree of attainable anisotropy. However, the solar wind is also a turbulent medium and thus the relevance of uniform equilibrium linear Vlasov instabilities is unclear.

In plasma theoretically unstable to kinetic instabilities, there are observed enhancements in wave power \citep{BaleEA09,BourouaineEA10}. These signatures have recently been reproduced using elevated local turbulent fluctuations in Hybrid Vlasov-Maxwell simulations \citep{ServidioEA13}, without invoking microinstabilities. Plasma unstable to mirror and firehose instabilities is also significantly hotter than under typical stable conditions \citep{LiuEA06,MarucaEA11}. These regions contain the most intermittent signatures \citep{OsmanEA12b,ServidioEA13}, which have properties consistent with coherent structures dynamically generated by strong plasma turbulence \citep{GrecoEA08,GrecoEA09,OwensEA11}. The same structures are statistically associated with plasma heating \citep{ParasharEA09,OsmanEA11,OsmanEA12a,WanEA12,WuEA13} and increased temperature anisotropy \citep{ServidioEA12,KarimabadiEA13}. These are candidate magnetic reconnection sites \citep{ServidioEA11}, where the onset of reconnection could be affected by the presence of temperature anisotropy-driven instabilities \citep{MatteiniEA13}. All these studies suggest underlying physical relationships exist between plasma turbulence, heating mechanisms in the solar wind and the kinetic physics that emerges from, or leads to, the growth of these instabilities. There are observations indicating similar instabilities are relevant for other solar wind plasma particles, such as helium ions \citep{MarucaEA12} and electrons \citep{StverakEA08}. Theoretical work predicts these instabilities are relevant in other astrophysical plasmas, such as accretion disks \citep{SharmaEA07,RiquelmeEA12} and galaxy clusters \citep{SchekochihinEA10,KunzEA11}. Hence, understanding the relationship between turbulence and proton microinstabilities in the solar wind could have far reaching implications for different particle species and astrophysical plasmas. This Letter addresses these important issues by presenting novel observational results linking instability thresholds and kinetic effects to direct estimates of the turbulent energy cascade rates.  

\textit{Analysis}.---We use around $4.5\times 10^{6}$ independent plasma and magnetic field measurements from the Wind spacecraft during the interval 1 Jan. 1995 to 31 Dec. 2011. The Faraday cup instrument in the Solar Wind Experiment (SWE) \citep{OgilvieEA95} measures 92 s resolution proton number density $n_{p}$, bulk velocity $\mathbf{v}_{sw}$, and proton temperature. This is separated into parallel $T_{\parallel}$ and perpendicular $T_{\perp}$ temperatures by comparison with the local magnetic field from the Magnetic Field Investigation \citep{LeppingEA95}. Only solar wind data is used, and measurements either in the magnetosphere or contaminated by terrestrial foreshock are removed. We also require the uncertainties in the plasma measurements to be less than 10\%. 

\begin{figure*}[t]
\includegraphics[width=17cm]{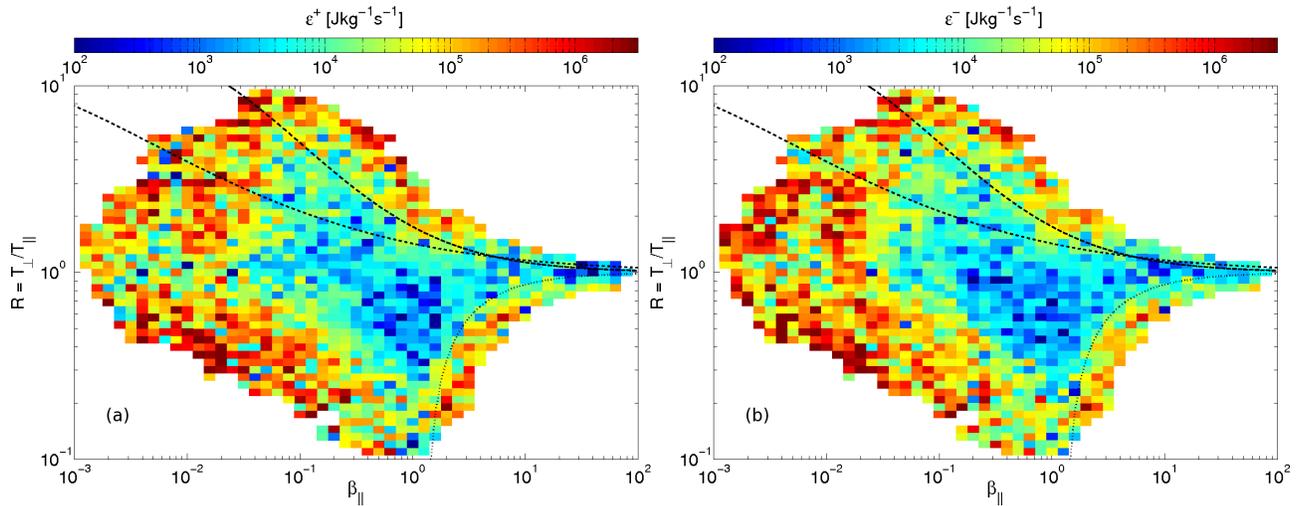}
\caption{Plot of the mean (a) antisunward and (b) sunward turbulence cascade rate over the $\left( \beta_{\parallel},T_{\perp}/T_{\parallel} \right)$-plane. 
The curves indicate constant values of
theoretical growth rates for the mirror (dashed), cyclotron (dot-dashed), and oblique firehose (dotted) instabilities. In both cases the energy cascade rate is significantly enhanced along the boundary of the $\left( \beta_{\parallel},T_{\perp}/T_{\parallel} \right)$-plane.}
\label{Fig:cr_comp}
\end{figure*}

Here we ask if kinetic effects, such as plasma heating and temperature anisotropy, are related to the turbulent character of the solar wind. To this end, an extension of the Kolmogorov-Yaglom law \citep{Kolmogorov41, Yaglom75} to time-stationary incompressible homogeneous magnetohydrodynamic (MHD) turbulence \citep{PolitanoPouquet98a,PolitanoPouquet98b} is used to calculate the energy cascade rate. This manifests as two symmetric scaling laws in terms of Els\"asser variables $\bf{z}^{\pm} = \bf{v} \pm \bf{b}$:
\begin{equation} 
\langle (\hat{\mathbf{r}}\cdot\delta\mathbf{z}^{\mp})\left|\delta\mathbf{z}^{\pm}\right|^{2}\rangle = -\frac{4}{3}\epsilon^{\pm}\left|\mathbf{r}\right|
\end{equation}
where the magnetic field fluctuations are normalized to Alfv\'en velocity units $\mathbf{b}/\sqrt{\mu_{0}m_{p}n_{p}}$, $\delta\mathbf{z}^{\pm} = \mathbf{z}^{\pm}(t + \delta t) - \mathbf{z}^{\pm}(t)$ are increments of the Els\"asser fields using a time lag $\delta t$, and $\epsilon^{\pm}$ are the respective energy cascade rates. The Els\"asser variables have been sector rectified such that $\mathbf{z}^{-}$ is sunward and $\mathbf{z}^{+}$ is antisunward. An ensemble average is denoted by $\langle\ldots\rangle$ and large amounts of data are usually required for statistical accuracy as the mixed third-order increment distribution is only slightly skewed and thus involves a lot of cancellation. However, the methodology employed here adopts a different approach based on conditional averages of the third-order law. 

This study is concerned with how cascade rates are distributed on the $(\beta_{\parallel}$,R)-plane rather than their actual absolute values, and we seek to maximize data coverage. Hence, cascade rates obtained without any averaging are compared to those obtained using averages of about 40 min to 6.7 h in duration, which corresponds roughly to 1--10 correlation times \citep{MatthaeusEA05}. Following binning and averaging in the $(\beta_{\parallel}$,R)-plane, all of these conditionally averaged cascade rate estimates displayed similar behavior, although the absolute values did differ. The rates were computed for several time lags $\delta t = \{92, 184, 368, 736\}$ s, which all correspond to spatial separations ($\mathbf{r} = -\mathbf{v}_{sw}\delta t$) within the inertial range when using Taylor's hypothesis \citep{Taylor38}. The cascade rates are independent of spatial separation in Eq. (1), which provides a further check on the reliability of these estimates. For each spatial lag, the behavior of the estimated cascade rates on the ($\beta_{\parallel}$,R)-plane is almost identical, as is consistent with theoretical expectation that $\epsilon^{\pm}$ is constant in the inertial range. Therefore, we use the single-point cascade rates in the proceeding analysis, supported by compelling evidence that the resulting behavior is robust. The globally averaged cascade and heating rates are $\epsilon^{+} = 4.6 \pm 0.3 \times 10^{3}$ Jkg$^{-1}$s$^{-1}$ antisunward and $\epsilon^{-} = 2.5 \pm 0.2 \times 10^{3}$ Jkg$^{-1}$s$^{-1}$ sunward, which is consistent with previous results \citep{MacBrideEA08,OsmanEA11b}. While these represent a forward (direct) cascade to smaller scales, the local cascade rate can occasionally suggest a dominant back-transfer (back-scatter). However, this is most likely a reflection of the inherent variability in the measurement rather than a physical feature, and so only the magnitude of the cascade rates will be considered (hereafter $\epsilon^{\pm}$ will refer to the cascade rate magnitude).   

\begin{figure}[h]
\includegraphics[width=8.5cm]{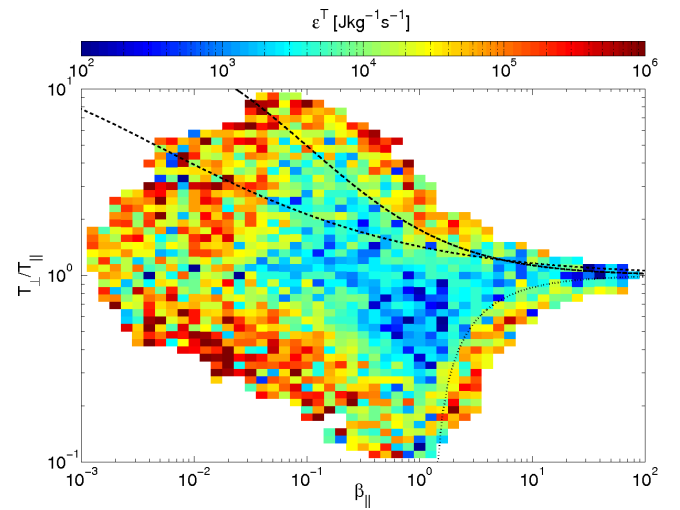}
\caption{Plot of the mean total turbulence cascade rate in the $\left( \beta_{\parallel},T_{\perp}/T_{\parallel} \right)$-plane. The curves indicate theoretical growth rates for the mirror (dashed), cyclotron (dot-dashed), and oblique firehose (dotted) instabilities. The total energy cascade rate is significantly enhanced along the boundaries.}
\label{Fig:cr_tot}
\end{figure}

\textit{Results}.---To generate Figs. 1--4 in this Letter, the selected observations were divided into a $50 \times 50$ grid of logarithmically spaced bins in the ($\beta_{\parallel}$,R)-plane. Any bins containing fewer than 50 records were discarded. Each plot in Fig. 1 and 2 is a composite of the listed parameter computed using four different spatial lags $\mathbf{r} = -\mathbf{v}_{sw}\delta t: \delta t = \{92, 184, 368, 736\}$ s. This approach provides a compromise between reasonable statistical accuracy and coverage over the ($\beta_{\parallel}$,R)-plane. 

Figure 1 shows the mean antisunward and sunward Els\"asser energy cascade rates in each grid-square within the ($\beta_{\parallel}$,R)-plane. While the noise is an expected consequence of our methodology, a clear pattern does emerge where most of the highest cascade rates are found along the boundaries and the lowest rates are mainly concentrated in the center. This suggests nonlinear couplings as measured by third-order statistics are strongest in the parameter regimes along the boundaries. The lower cascade rates are consistent with previous estimates \citep{MacBrideEA08,OsmanEA11b}, and thus it is reassuring to find them located in the ($\beta_{\parallel}$,R)-plane region associated with the greatest data population \citep{BaleEA09}. The observed behavior is very similar for both $\epsilon^{+}$ and $\epsilon^{-}$. These results are supported by Fig. 2, which shows the total cascade rate $\epsilon^{T} = (\epsilon^{+} + \epsilon_{-})/2$ is enhanced in the peripheries of the ($\beta_{\parallel}$,R)-plane. While these regions have the least data, the robust trend of elevated cascade rates cannot solely be a consequence of noise.

\begin{figure}[h]
\includegraphics[width=8.5cm]{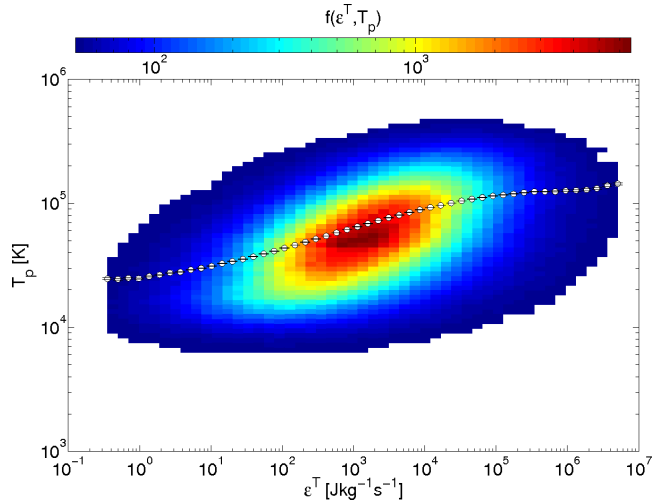}
\caption{Joint histogram between total turbulence cascade rate and proton temperature, where the mean $T_{p}$ in each cascade rate bin is represented by a white dot. There is a clear correlation between enhanced $T_{p}$ and high energy cascade rates.} 
\label{Fig:tempHist}
\end{figure}

The distribution of cascade rates in Figs. 1--2 have statistical significance and suggest implications for the relationship between turbulence, kinetic effects and instabilities. However, the solar wind does not populate the ($\beta_{\parallel}$,R)-plane uniformly and this must be considered in order to correctly interpret the results. Therefore, it is instructive to examine the joint distribution between total cascade rate and proton temperature $f(\epsilon^{T},T_{p})$ as shown in Fig. 3 for the entire solar wind dataset. Here it is immediately clear that the bulk population is located roughly in the center. It is also apparent that the distribution is slanted upward, suggesting a correlation between $\epsilon^{T}$ and $T_{p}$. This is more manifest when $T_{p}$ is averaged in contiguous $\epsilon^{T}$ bins. Therefore, regions of elevated cascade rate are statistically linked to enhanced proton temperatures. Indeed, the pattern of high cascade rates in Figs. 1--2 closely resembles the distribution of increased temperatures at a given $\beta_{\parallel}$ observed by \citep{MarucaEA11}.    

Recent plasma simulation results have implied that kinetic effects, such as proton \citep{ServidioEA12,KarimabadiEA13} and electron \citep{HaynesEA13} temperature anisotropies and other departures from isotropic Maxwellian distributions \citep{GrecoEA12}, are associated with intermittent plasma turbulence. In a complimentary study \citep{ServidioEA13}, temperature anisotropy appears to be a consequence of large turbulent fluctuations and it is this dynamical generation of anisotropy that allows the solar wind to populate different regions of the ($\beta_{\parallel}$,R)-plane. Hence, it is instructive to determine whether proton temperature anisotropy is correlated with the estimated cascade rates. Figure 4 shows the joint distribution $f(\epsilon^{T},T_{L}/T_{S})$, where $T_{L}$ ($T_{S}$) is the largest (smallest) of $T_{\perp}$ and $T_{\parallel}$. The distribution is skewed and the averaged $T_{L}/T_{S}$ in adjacent $\epsilon^{T}$ bins clearly demonstrates a positive correlation. This is also shown in Figs. 1--2, where the highest rates are mostly in regions away from isotropy. 

\begin{figure}[t]
\includegraphics[width=8.5cm]{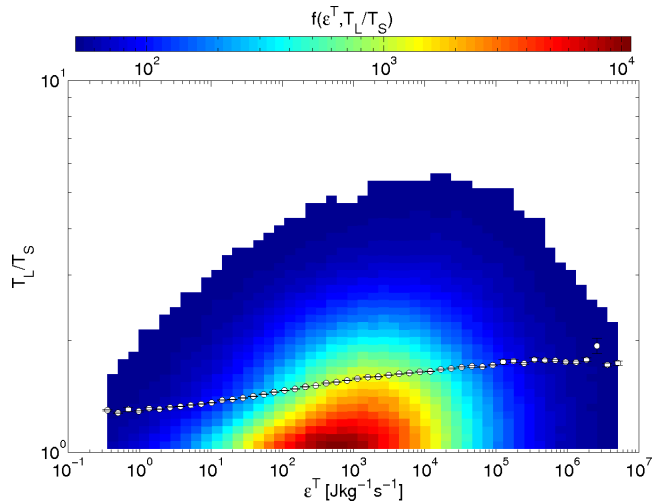}
\caption{Joint histogram between total turbulence cascade rate and proton temperature anisotropy, where $T_{L}$ ($T_{S}$) is the largest (smallest) of $T_{\perp}$ and $T_{\parallel}$. The mean temperature anisotropy in each cascade rate bin is represented by a white dot. The highest energy cascade rates are associated with the greatest average departures from temperature isotropy.} 
\label{Fig:aniHist}
\end{figure}

\textit{Discussion}.---We have presented the first direct evidence that inertial range turbulent cascade rates in the solar wind are statistically associated with proton kinetic effects at the extremes of the ($\beta_{\parallel}$,R)-plane. These regions have previously been linked with helium and electron temperature anisotropies, intense coherent structures, enhanced turbulent fluctuations, and linear Vlasov instability thresholds. It has until now been possible to interpret these observed statistical correlations as consequences of either plasma turbulence or microinstabilities. However, the turbulent cascade rate has no correspondence in linear Vlasov theory and so enhancements cannot be caused by instabilities. In the nonlinear regime, instabilities such as the parallel firehose \citep{RosinEA11} can drive a self-similar inverse cascade and there are indications of energy injection into parallel wavenumbers around the ion Larmor scale \citep{WicksEA10}. However, this energy injection appears to be localized in wavenumber and the observed power spectrum scaling is inconsistent with predictions of a cascade from instabilities. Therefore, we conclude that the current evidence favors an interpretation in which a turbulent cascade from inertial to kinetic scales is the causal agent allowing the solar wind to populate the extremes of the ($\beta_{\parallel}$,R)-plane. This is consistent with recent Vlasov simulations that find elevated turbulent fluctuations generate increased temperature anisotropy \citep{ServidioEA12,ServidioEA13}.

While instabilities may act to confine the solar wind plasma, turbulent fluctuations and cascade rates can cause temperature anisotropies, intermittent structures and heating in the ($\beta_{\parallel}$,R)-plane. These kinetic effects, especially elevated $T_{\perp}$, have all previously been linked to resonant cyclotron damping \citep{Marsch06}. Indeed, a recent detailed analysis \citep{KasperEA13,MoyaEA13} has revealed ion cyclotron resonance contributes to heating the solar wind, and is linked to differential flow and temperature anisotropies. If cyclotron damping and instabilities can occur in response to plasma turbulence, then a consistent explanation of the present, as well as previous, results would emerge. Hence, it is encouraging that strong cyclotron interactions of the betatron type have been found in studies of test particles in weakly three-dimensional and highly anisotropic magnetohydrodynamic turbulence \citep{DalenaEA13}.  

This research is supported by UK STFC, EU Turboplasmas project (Marie Curie FP7 PIRSES-2010-269297), NASA Magnetospheric Multi-Scale Mission Theory and Modeling program (NNX08AT76G) and Solar Probe Plus ISIS project, and NSF SHINE (AGS-1156094) and Solar Terrestrial (AGS-1063439) programs.

\end{document}